\documentclass{llncs}

\usepackage[utf8]{inputenc}
\usepackage{amsmath}
\usepackage{graphicx}
\usepackage{amsfonts}
\usepackage{amssymb}
\usepackage{url}

\usepackage{caption}
\usepackage{subcaption}
\usepackage[misc,geometry]{ifsym}
\graphicspath{ {./images/} }

\begin{document}

{\let\thefootnote\relax\footnotetext{Copyright \textcopyright\ 2020 for this paper by its authors. Use permitted under Creative Commons License Attribution 4.0 International (CC BY 4.0). CLEF 2020, 22-25 September 2020, Thessaloniki, Greece.}}

\title{Revealing Lung Affections from CTs.\\ A Comparative Analysis of Various Deep Learning Approaches for Dealing with Volumetric Data}

\author{Radu Miron\inst{1,2}
, Cosmin Moisii\inst{1,2}
, Mihaela Elena Breaban \Letter
\inst{1,2} 
}

\institute{
SenticLab, Iasi, Romania\\
\and
Faculty of Computer Science, "Alexandru Ioan Cuza" University of Iasi, Romania\\
\url{pmihaela@info.uaic.ro} 
}

\maketitle

\begin{abstract}
The paper presents and comparatively analyses several deep learning approaches to  automatically  detect  tuberculosis related lesions in lung CTs, in the context of the ImageClef 2020 Tuberculosis  task. Three classes of methods, different with respect to the way the volumetric data is given as input to neural network-based classifiers are discussed and evaluated. All these come  with a rich experimental analysis comprising  a  variety  of  neural  network  architectures,  various  segmentation  algorithms and data augmentation schemes. The reported work belongs to the SenticLab.UAIC team, which obtained the best results in the competition.
\end{abstract}

\section{Introduction}
Medical imaging technologies like Computer Tomography (CT) and Magnetic Resonance (MR) produce high volumes of data in the form of volumetric images. The richness of information they provide is essential to correct diagnosis but brings at the same time new challenges, both for manual/human and automatic/machine processing: these are not only about the size of the produced data but also about the complexity of the diagnosis process itself. With respect to automated diagnosis, the volumetric images, which can be seen both as matrices of pixels/voxels or series of 2D images (usually called slices), produced high effervescence in the deep learning research community, triggering a variety of new architectures and approaches. 

The current paper makes use of deep learning to automatically detect tuberculosis and related affections in lung CTs, in the context of the ImageClef Tuberculosis task \cite{ImageCLEFTBoverview2020,ImageCLEF20}. We investigate three types of approaches, different with respect to the way the volumetric data is given as input to neural network-based classifiers. One type, popular among the participants in the previous year competition \cite{cid2019overview}, is based on reducing the volumetric image to a small set of 2D projections. Obviously, this approach consistently reduces the size of the data to be processed by the classifier but inherently may lose important information. The second type exploits the whole data matrix by using 3D convolutions or by fusing the information from the slices. The third type, which was ranked as the winner of the 2020 evaluation session, consists in moving the decision layer from the whole volume of data to the slice level. All these three different approaches come with a variety of neural network architectures, various segmentation algorithms and data augmentation schemes. The work reported stays behind the SenticLab.UAIC team, obtaining the best results in the competition\footnote{https://www.imageclef.org/2020/medical/tuberculosis/} \cite{ImageCLEFTBoverview2020}.

The paper is structured as follows. Section \ref{sec:data} describes the challenge and the dataset. Section \ref{sec:proj} describes the approaches developed based on reducing the volumetric image to 2D projections, starting with the previous year winning approach reported in  \cite{liauchuk2019imageclef}, which we further enhanced to address the 2020 tasks. Section \ref{sec:whole} presents the approaches we used to exploit the whole volumetric information. Section \ref{sec:slice} describes the architectures used to process the information at slice level and the heuristics used to produce the diagnosis report at the CT level. Because of the large number of approaches we evaluated, some of them abandoned earlier (not submitted in the competition) due to poor results on our local validation data, we report and discuss performance results immediately after each method description.  Section \ref{sec:results} summarises the results for the best approaches that were evaluated on the blind test set in the competition and discusses comparatively the performance of the three classes of methods. Section \ref{sec:conclusions} concludes the paper.

\section{ImageClef Tuberculosis: tasks, data, evaluation}\label{sec:data}
The challenge in the 2020 ImageClef Tuberculosis competition is the automatic detection of tuberculosis and related lesion types in CTs. The CT report to be generated must contain 3 binary labels for each lung, indicating the presence of TB lesions in general, the presence of pleurisy and caverns in particular.

The training dataset consists of 283 CTs. All CTs present at least one lung affected, 19 have pleurisy and 126 caverns. Because we split each CT into left/right lungs, this translates to 566 inputs, 444 affected, 21 with pleurisy and 145 with caverns.

The task is therefore a multi-binary classification problem, with three target labels per lung. For each target label the AUC is computed and the ranking is done on a test set, first by computing the average AUC and then by the minimum AUC over the 3 target labels.

We split the data into train/validation in the same fashion as \cite{liauchuk2019imageclef} setting apart every 4th input into the validation set and use this configuration throughout the competition.

\section{Squeezing Volumetric Data: 2D Projections}\label{sec:proj}
The 3D matrix representing the volumetric image can be reduced to simpler 2D representations by traversing it in each of its three dimensions and computing statistics on numeric vectors. In the case of lungs CTs, a segmentation algorithm is firstly applied to detect the lungs and eliminate the other parts in the CT. Further, we used the method proposed in \cite{liauchuk2019imageclef}, where the mean, the maximum and the standard deviation is computed on each direction, generating three 2D matrices which can be interpreted as an RGB image (a single 2D image with 3 channels). All the processing steps described in \cite{liauchuk2019imageclef} are kept: mask erosion,  increasing the voxels intensity in the CT by 1024HU, dividing the mean values and standard deviations values (red and blue channels) by their maximum, dividing the maximum values (green channel) by 1500. At the end, for each lung we have a set of three 2D RGB images, each image corresponding to one of the three dimensions of the 3D matrix.

\subsection{The Impact of Segmentation}
\begin{figure}[h]
\includegraphics[scale=0.25]{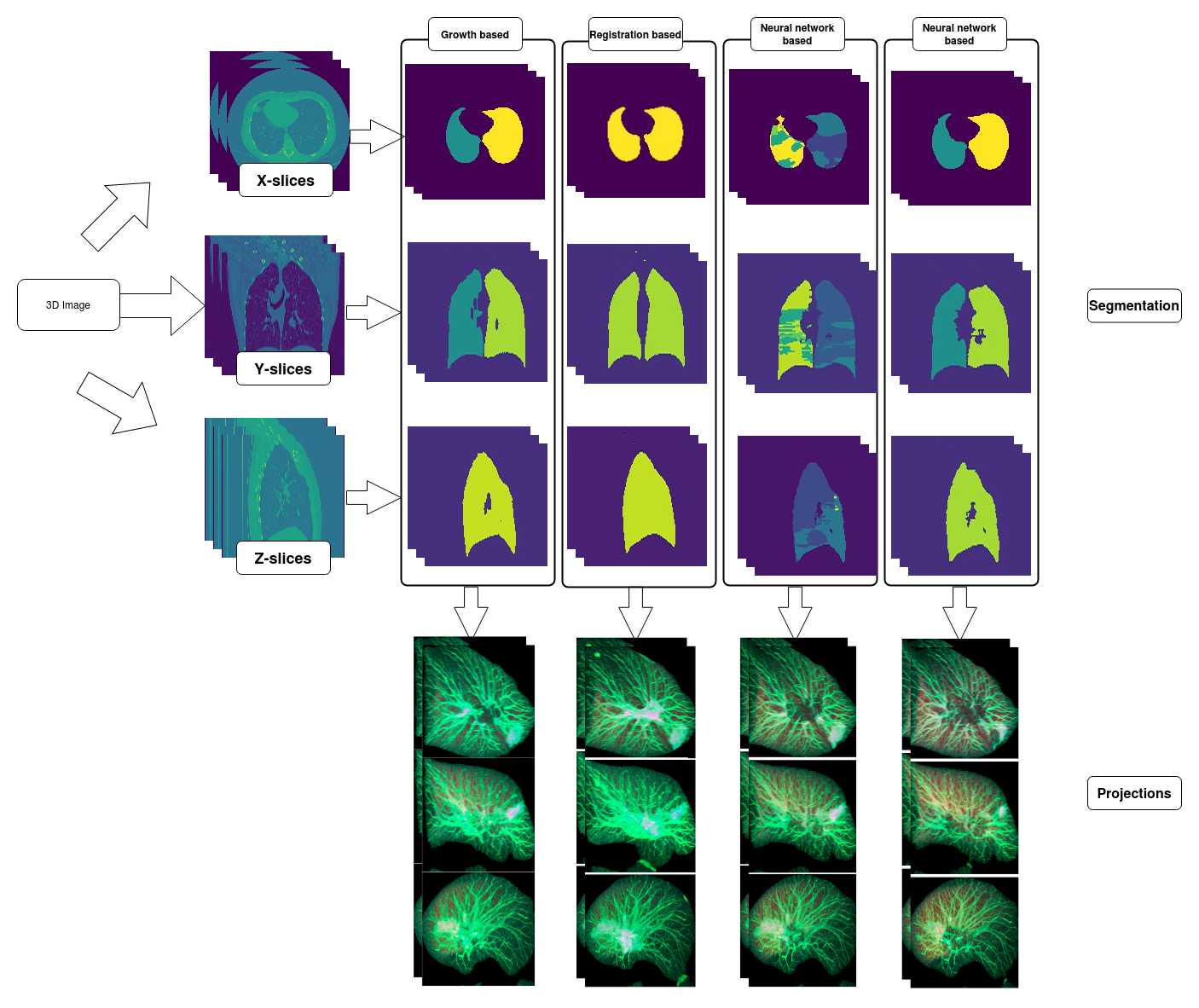}
\caption{Projections dataset creation flow using 4 segmentation variants (in order: growth-based, registration-based, unetLTRCLobes and unetR231). Although the 4 types of projections resulted look only slightly different, the difference in classification scores is significant.}
\label{fig:}
\end{figure}
The first step behind all our approaches is image segmentation, with the aim of identifying and isolating each lung in the volumetric image. Because the performance of further processing is greatly influenced by the quality of segmentation (especially in the case of the 2D projection approach where the projections take into account the entire volume), we tested several segmentation methods. The organizers provided for all patients two versions of automatically extracted masks of the lungs: one which relies only on anatomical assumptions \cite{segm_fully_automatic_multistage}, and one based on non-rigid registration \cite{segm_nonrigid_reg}. Additionally, we used \emph{U-net(R231)} and \emph{U-net(LTRCLobes)}  which were pre-trained for lung segmentation on large and diverse datasets \cite{segm_Unet} \footnote{https://github.com/JoHof/lungmask}. Our experiments show that the  first technique based on anatomical assumptions behaves much like region growing not being able to catch holes or necrotic tissue in lungs, the second technique manages to capture necrotic tissue while the ones based on U-net include airpockets, tumors and effusions. The flow of the dataset creation together with some projections corresponding to several segmentation techniques can be seen in fig. 1.

Feeding a VGG neural network \cite{VGG} with 2D projections obtained on the segmented volumetric image, the average AUC scores obtained on our validation set indicate the registration based method to give the best performance (\textbf{AUC=0.693}) followed at small distance by \emph{U-net(R231)} (\textbf{AUC=0.674}) and \emph{U-net(LTRCLobes)} (\textbf{AUC=0.668}), but consistently surpassing the anatomy-based method (\textbf{AUC=0.580}).

Consequently, all our further experiments use the segmentation provided by non-rigid registration \cite{segm_nonrigid_reg}.

\subsection{Data Augmentation}
The images go through a series of augmentations, each with a certain probability of being applied, including: horizontal and vertical flipping, small degrees of rotations, blurring, added gaussian noise, distortions, random cropping, and changing different values of hue, saturation or brightness. We used the Albumentations \footnote{https://github.com/albumentations-team/albumentations} library for most of these augmentations.

\subsection{The 2D Approach with Preprocessing (\emph{PreProcProj})}
\begin{figure}
\centering
\includegraphics[scale=0.151]{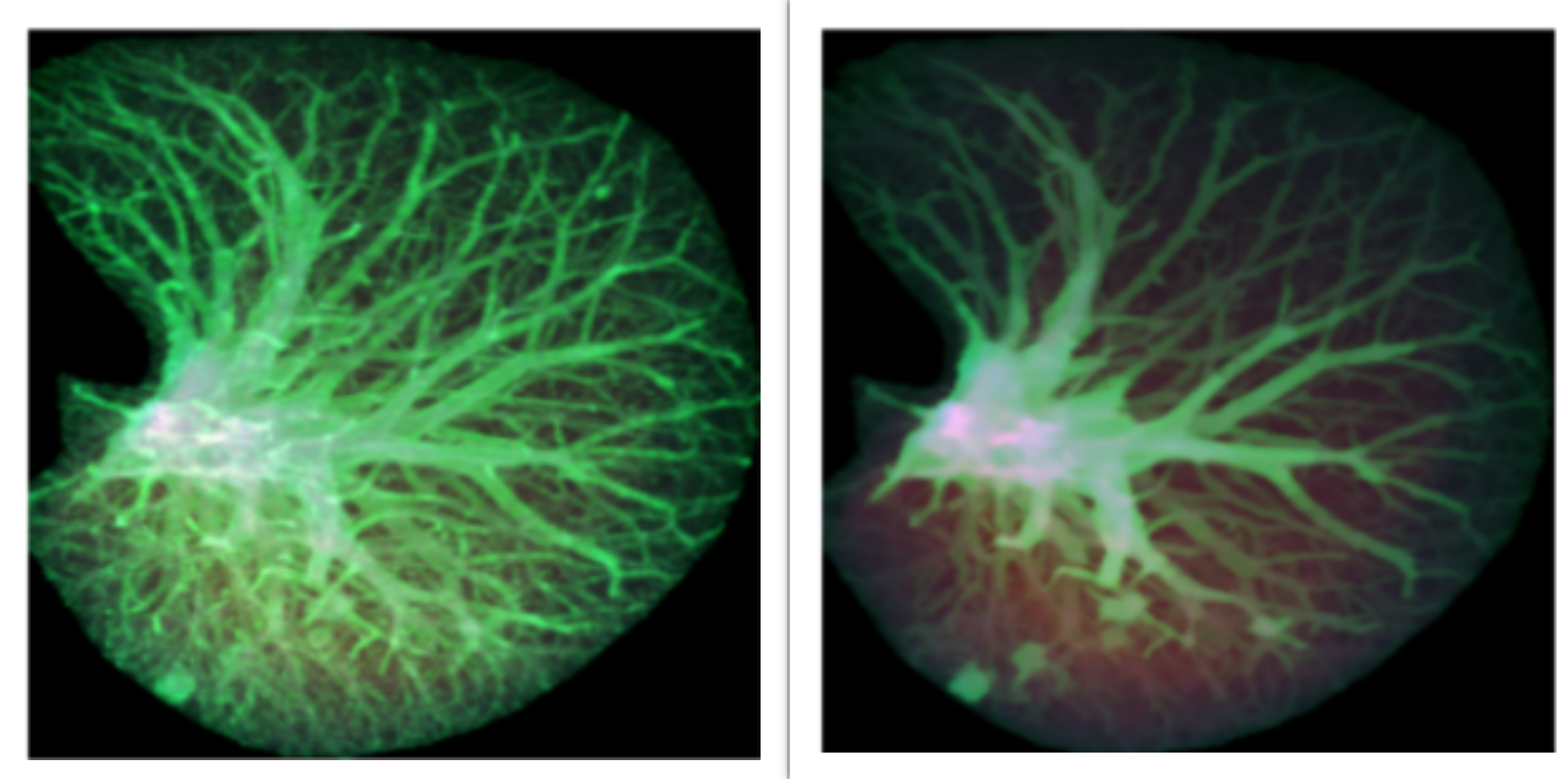}
\caption{Comparison between projection without(left) and with pre-processing(right)}
\label{fig:vessels}

\end{figure}

In an effort to improve over the last year result, we used the pre-processing provided in \cite{perez2017automated}, with the aim to eliminate the small vessels from the projection, thus making the affected area more obvious. We adopted all the pre-processing steps that the authors mention, except the regional maxima calculation. Figure \ref{fig:vessels} illustrates the difference between projections with and without further pre-processing. 

For training we chose AlexNet\cite{krizhevsky2012imagenet}. The input consists of the three projections of a volume, on each axis. After extracting features from each projection with AlexNet, we concatenate all the features, feed them into a linear layer and predict probabilities for a lung to have affections, caverns, pleurisy or be healthy. 
With this approach we scored an \textit{AUC} of 0.793 on the test set.

\subsection{The 2D Approach Scoring the Best (\emph{ResNet50Proj})}

We further tried different variants of Resnet\cite{he2016deep} 
and SqueezeNet\cite{iandola2016squeezenet}. 

\begin{figure}[t]
\includegraphics[scale=0.25]{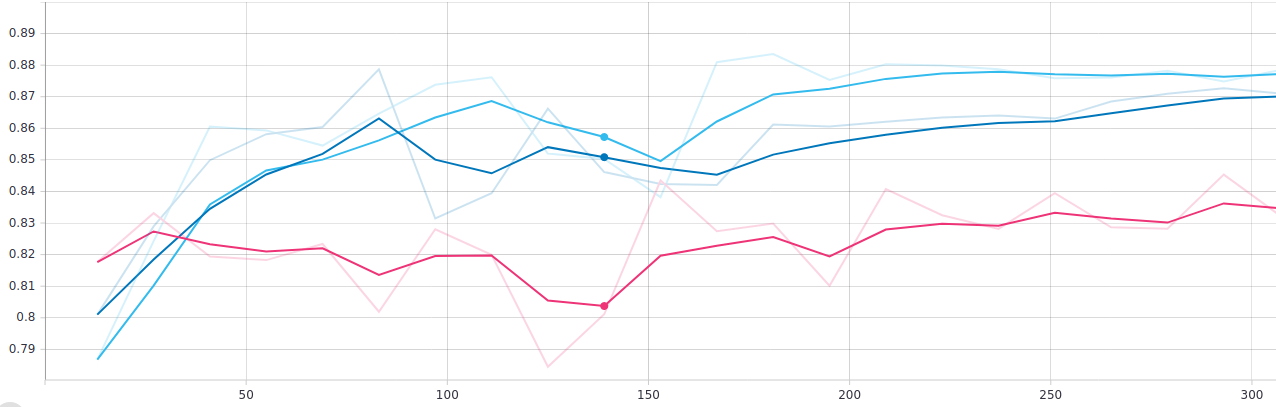}
\caption{The AUC score progress on the hold-out validation set for different resnet models. Pink: resnet34 with no augmentations, DarkBlue: resnet34 with augmentations, LightBlue: resnet50 with augmentations}
\label{fig:AUCscores}
\end{figure}

We extracted the lungs using the registration-based segmentations, computed all 3 projections, split by lung side and processed with the augmentation we listed in section 3.2. We trained the networks and aggregated the results on all 3 projections and computed the mean score to obtain the final results back at CT level. We tried different approaches in aggregating the results including training a small neural network, but found the simpler mean aggregation to give the highest score. We thus obtained our highest score in the 2D approach using a resnet-50 network \cite{he2016deep} pretrained on Imagenet\cite{deng2009imagenet} with an AUC on our hold-out validation set of 0.877; however this result was not submitted. Our first submission to the competition was a resnet34 model with no augmentations which obtained on the hold-out validation set and on the test set the same AUC score of 0.825. In Figure \ref{fig:AUCscores} we can see the \textit{AUC} progress on different models we tried.

\section{Exploiting Volumetric Data as a Whole}\label{sec:whole}
\subsection{3D Convolutions}
In an attempt to make use of the whole volume at once, we used SqueezeNet in a 3D version, based on the implementation found in the repository \footnote{https://github.com/okankop/Efficient-3DCNNs}. In order to work with volumes of different sizes, we used batch size equal to 1. To handle volumes with a large number of slices, we used Apex\footnote{https://github.com/NVIDIA/apex} library for reducing the burden on our GPU. In a preliminary experiment we considered only the case \textit{affected vs. not affected}. We noticed the bad results during the training: after some epochs the prediction scores stagnated, for all volumes, between $0.4$ and $0.6$. We concluded that 3D convolutions are not able to capture the important information on our small training set of volumetric images. 

\subsection{Slices fusion}
In our attempt to associate the entire volumetric image to a label, we constructed a hybrid approach. We fed the volume slice by slice into a convolutional neural network, fused the resulted feature maps at channel level and continued with another small convolutional network into a prediction.
The initial convolutional neural network is composed from the encoder part of a U-net\cite{ronneberger2015u} architecture which was pretrained on a segmentation task at the end of which we applied a squeeze connection to reduce the number of channels, fused the resulting feature maps so that the slices processed in parallel by the CNN would now be treated as channels of a single input, then applied a resnet-like small network to compile the features into a label.
We no longer use the masks to extract the lungs but instead use a simple threshold based segmentation to compute the boundaries of the body and crop out the space around the it. We again split by lung side and used only horizontal flip as a preprocessing. The resulting volume is resized to the fixed size of (128, 256, 256) The network could then be fed images in batches multiple of 128 representing the slices of a volume.

To make maximum use of the GPU memory, we used the Apex library to train using mixed precision, in a distributed manner on 2 GPUs. We could fit 2 times 128  images into the memory corresponding to 2 volumes. 

The approach turned out to be cumbersome. The time to process an epoch was relatively high and the convergence of the network seemed slow. After 2 days of training we decided to stop and the network reached an AUC of around 0.6 on the hold-out validation set.

\section{Sequencing Volumetric Data: a Slice by Slice Classification Approach}\label{sec:slice}

Having a closer look at the training set, one can observe that usually the lesions on the lungs are located only on a small number of slices from the whole volume. A natural idea is to try a 2D model that could differentiate between healthy lung slices and lung slices with lesions (caverns, pleurisy and affections) and construct the CT report based on the findings at slice level. For this purpose we need training data labeled at slice level and not CT level.

The first approach was to try to automatically detect the slices presenting lesions in the training set, using a lung nodule detector\footnote{https://github.com/BCV-Uniandes/LungCancerDiagnosis-pytorch} constructed by the winners of a challenge in cancerous nodules detection. The results were bad, the model not being able to recognize the slices showing caverns although these correspond to big, obvious regions.

Therefore, we started to manually select from each volume of the training set the slices with lesions. We actually found that this was not as time-consuming as we initially thought, by processing only the volumes labeled with lesions, and it definitely was worth the effort, as the increase in performance shows. 
The caverns are usually big and obvious and the affections are either nodules or dusty lungs images (which may indicate pneumonia), with very rare cases of pneumotorax (the lung disappearing due to the outbreak of a cavern). There are many cases when these lesions appear on a very small number of slices and thus, the two approaches described in sections \ref{sec:proj} and \ref{sec:whole} might have not been able to reveal them. 

\subsection{InceptionNet}
In our first tries using the slice by slice approach, we used \textit{InceptionNet
version 3}\cite{szegedy2016cvpr}. We used the annotated data in different ways. 
Transforming each slice of a volume into a picture, resizing each image to a size of $299 \times 299$, 
cutting the picture in half to obtain the two lungs and using vertical 
flip for all the pictures which are either affected, with caverns or with pleurisy, are
the data pre-processing steps for our first attempt using this approach. This
approach does not use the provided segmentation masks at all. We only used 4 labels as output: \textit{affected, caverns, ok, pleurisy}. 

\begin{figure}
\centering
\includegraphics[scale=0.151]{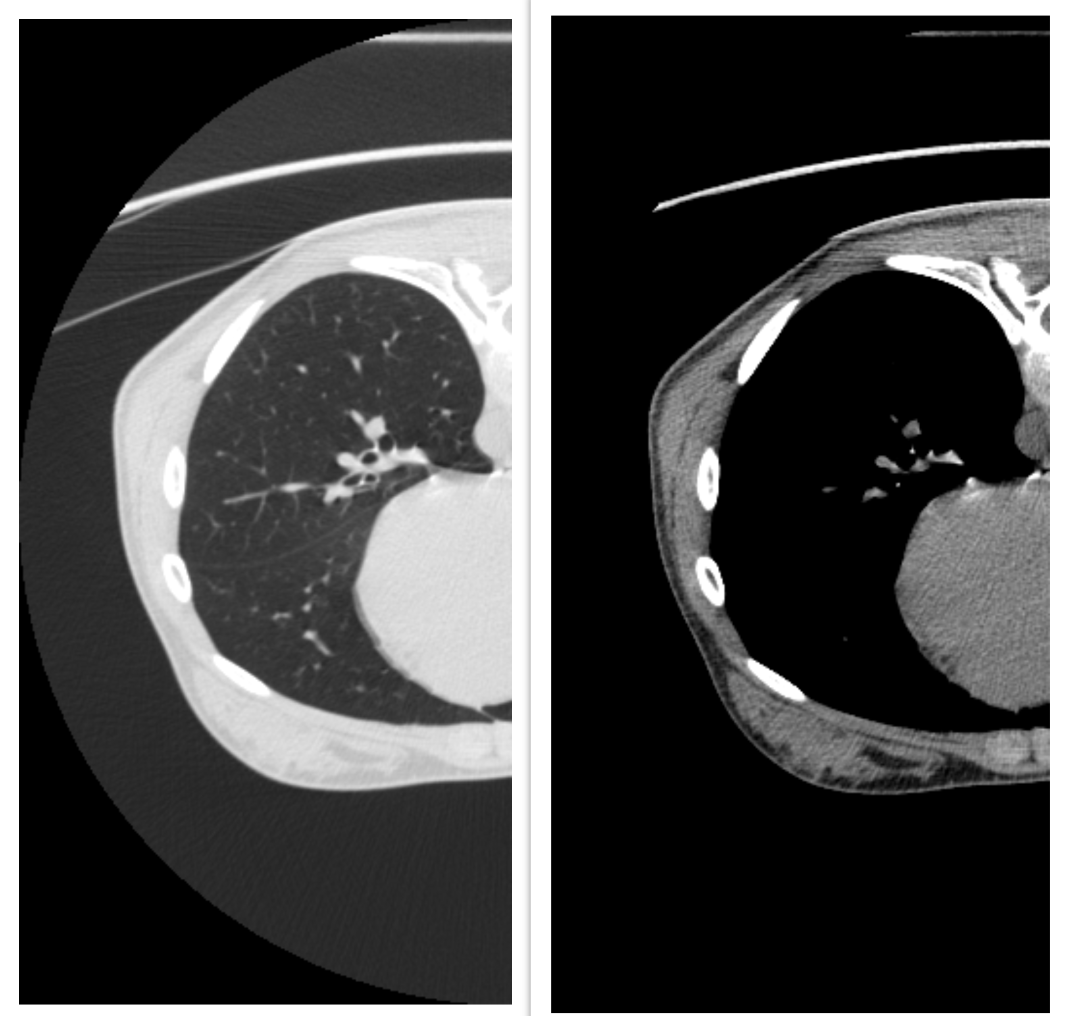}
\includegraphics[scale=0.151]{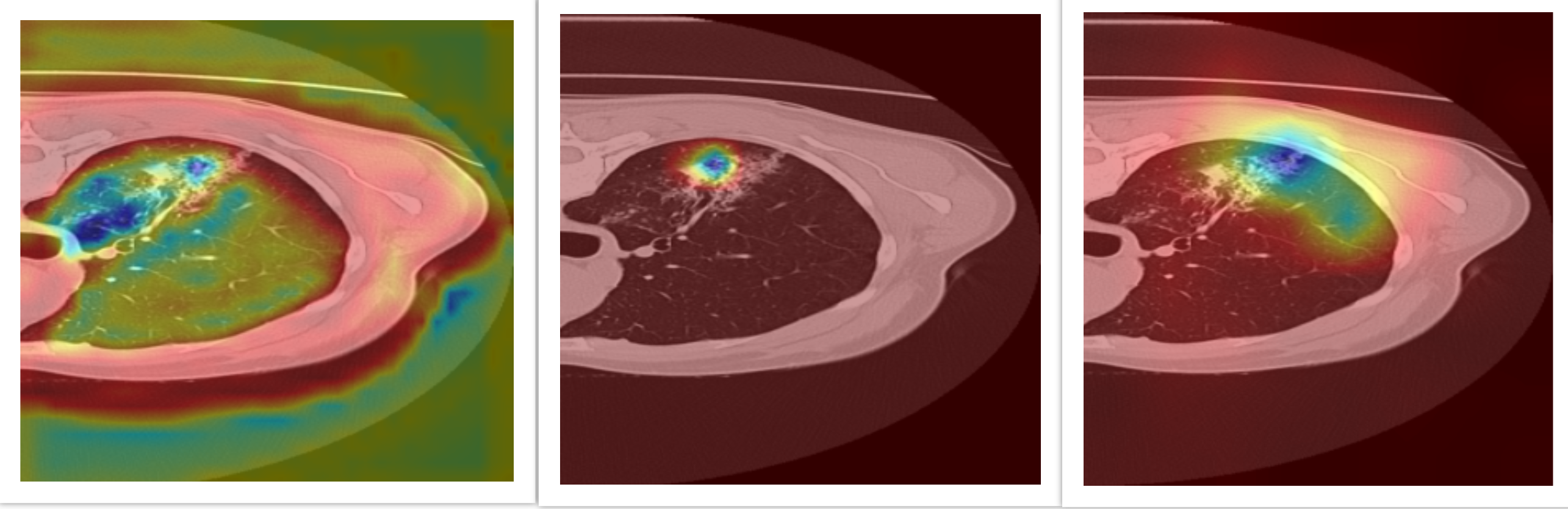}

\caption{Top: Comparison between different threshold values for the HU units\\ Bottom: Attention visualization for the three layers from InceptionNet}
\label{fig:attention}
\end{figure}

As input for the neural network we used several versions, having all images as 3-channel images:

a) \textit{NaiveInception}. We added a linear layer on top of the last adaptive average pooling
layer of the architecture, keeping the original InceptionNet weights freezed. With this approach we scored 0.86 \textit{AUC} score on the test set, surpassing this way the best approach based on 2D projections.

b)\textit{ThresholdInception}. The other approaches consist in using some other pre-processing steps. This time, when creating the photos from the 3D volume, we used
a Window Width and Window Level equal to 1500, -500 respectively. This way we improved the results to 0.887 mean \textit{AUC} and to 0.82 min \textit{AUC} on the test set.

c) \textit{TwoPicInception}. One other approach consists in mimicking the protocol a doctor has to follow
in order to decide affections(including caverns) and pleurisy. In order 
to see the affections more clearly, a doctor uses Window Width and Window Level equal to 1500, -500 
respectively, whereas for better visualisation of pleurisy, a doctor looks
at pictures with Window Width and Window Level equal to 350, 50 respectively. 
With this thresholding, the liquid surrounding the pleura becomes more observable. Figure \ref{fig:attention} top shows the differences between the two pictures.

In order to use information from 2 pictures during training, we used two InceptionNet 
modules, with trainable parameters and concatenated the two outputs of the 
adaptive average pooling layer. The decision was made based on the output of the last
linear layer applied on the concatenation discussed above. With this approach
we scored 0.89 mean \textit{AUC} on the test set.

d) \textit{AttentionInception}.We wanted to gain insight into how accurate the methods can be. We tried to check the predictions produced by the methods proposed and discovered that the models found
in a big proportion correct slices of the volumes which contained certain affections.
In order to work on the explainability of our model, we modified the structure of ThresholdInception,
using the idea from \cite{jetley2018learn}, introducing an attention mechanism. Instead
of feeding the output of the last pooling layer into the linear layer, we used
dot product attention \cite{bahdanau2014neural}. We computed similarity scores between the output
of the pooling layer and three different convolutional layers in the architecture.
After using the compatibility
scores as weights for the features extracted by the three layers, we concatenated the new
features. Using a linear layer on top, we predicted scores for the 4 categories.
After plotting the attention we noticed that the attention on the first layer selected
highlighted the whole lung area - supporting the idea that we don't need the segmentation masks, 
whereas the second layer of attention highlighted affections on the lungs. 
With this approach we scored 0.85 mean AUC. We believe this lower performance is due to the 
fact that the attention on the last layer was not good. Checking the visualisation for 
that layer we noticed useless areas highlighted (Fig. \ref{fig:attention}, bottom). Because of the limit imposed on the number of 
submissions, we stopped investigating this direction.

For all the methods above based on InceptionNet, training was performed for 30 epochs on one GPU Nvidia RTX 2070 with 8GB of memory, using Stochastic gradient descent optimizer. We divided the learning 
rate at each 10 epochs by 10 and used binary cross-entropy as loss function.

In order to establish the diagnosis for a volume we applied the following heuristic: we applied the inference step on all the pictures/slices from the volume and for each of the possible classes we took the maximum score encountered; if only one slice was found with an affection score higher than 0.8, then we divided the score of affected by 2.

\subsection{EfficientNet}
In an effort to use a powerful, yet small footprint network, in our last approaches we used efficientnet\cite{tan2019efficientnet}, specifically the b4 variant which has only 19M parameters but reaches top 1 accuracy of 82,6\% on Imagenet. We used a Pytorch implementation pre-trained on Imagenet\cite{deng2009imagenet}. 

The preprocessing we used here is similar to the ones we used before and took place at run-time on load. We applied the registration-based mask per slice based on a threshold to crop the body and remove much of the surrounding space, split the lungs into left/right (just by using splitting the image in half) and applied the same series of augmentations as in the previous approaches. The split and cropped image has dimension 256 $\times$ 256 and after randomly cropping it reduces to 224 $\times$ 224.
For ease of working we also kept the size of the volumetric image depth to a fixed 128 slices per volume.

If otherwise specified for this approaches as for the others we used a window level of -500 and range of 1500 corresponding to the most common values used in areas of acute differing attenuation values (example: lungs) where air and vessels will sit side by side.  

As input we tried several options, all of them maintaining 3 channels per image:

a) Micro-volumes (\emph{MicroVolSlice}): The importance of volumetric data is evident. This seemed especially apparent when we try to manually identify caverns which can present as rounded or irregularly shaped black centers surrounded by a white contoure. The caverns can range in size from small with a thin contoure line to large with thick and diffuse borders. The small caverns we found especially hard to identify as it can be confused with a section of a larger blood vessel. As untrained individuals, to eliminate the confusion we traced the potential cavern a few slices up or down to verify if it continues into a vessel or forms a pathology. To try and mitigate this type of confusion in a model we composed the 3 channels of the image from 3 consecutive (or equidistant) slices. In case the slice is at the beginning or end of the sequence we simply duplicated it to fill the channels.
\begin{figure}
\includegraphics[scale=0.21]{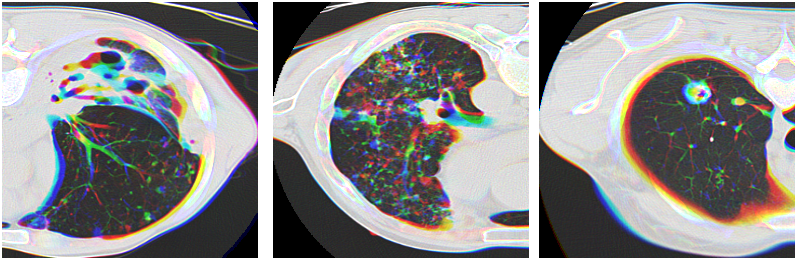}
\includegraphics[scale=0.21]{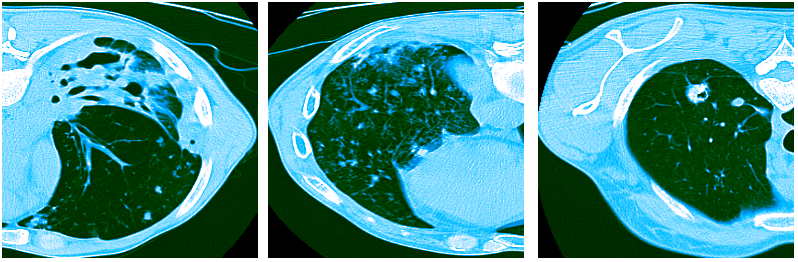}
\includegraphics[scale=0.21]{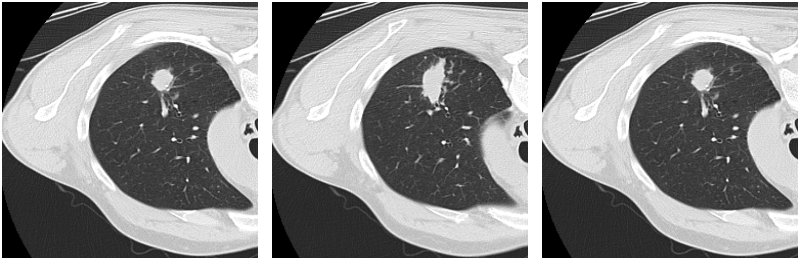}
\caption{Top left: sample of micro-volume images. Top right: Sample of false-color images. Bottom: Sample of "naive" images.}

\label{fig:}
\end{figure}

b) False-color (\emph{FalseColorlSlice}): To make use of the entire range of values of an MRI image we established 3 intervals in the Hounsfield units range to correspond to the 3 channels of an image. The window size and level used \footnote{https://radiopaedia.org/articles/windowing-ct} are (1500, -500) corresponding to the usual values used for lung imaging, (350, 40) called narrow window (used when examining areas of similar attenuation, for example, soft tissue) and (500, -600) a narrower window of the usual values for lung imaging in an attempt to retain more information around the values corresponding to blood vessels and soft tissues. The
result with this method however was not submitted to the site as the result on the hold-out set was poorer than the others.

c) Naive (\emph{NaivelSlice}): The image is simply duplicated into the 3 channels.


To compile the final results for each volume we aggregate the individual results per slice and choose the max of each each label across all the slides, which are then used to compute the AUC.

Loss: Cross entropy vs Binary cross entropy : A strange case comes from the fact that using the CrossEntropyLoss (on a multi-label classification) without softmax before the loss (thus assigning a predominant label for each slice) we obtained higher results than using BinaryCrossEntropyLoss.
The nature of the results is also very different, the first giving results on the extremes while the latter hovering around 0.5, but both giving decent results around $90\%$ AUC.

\begin{figure}[t]
\includegraphics[scale=0.25]{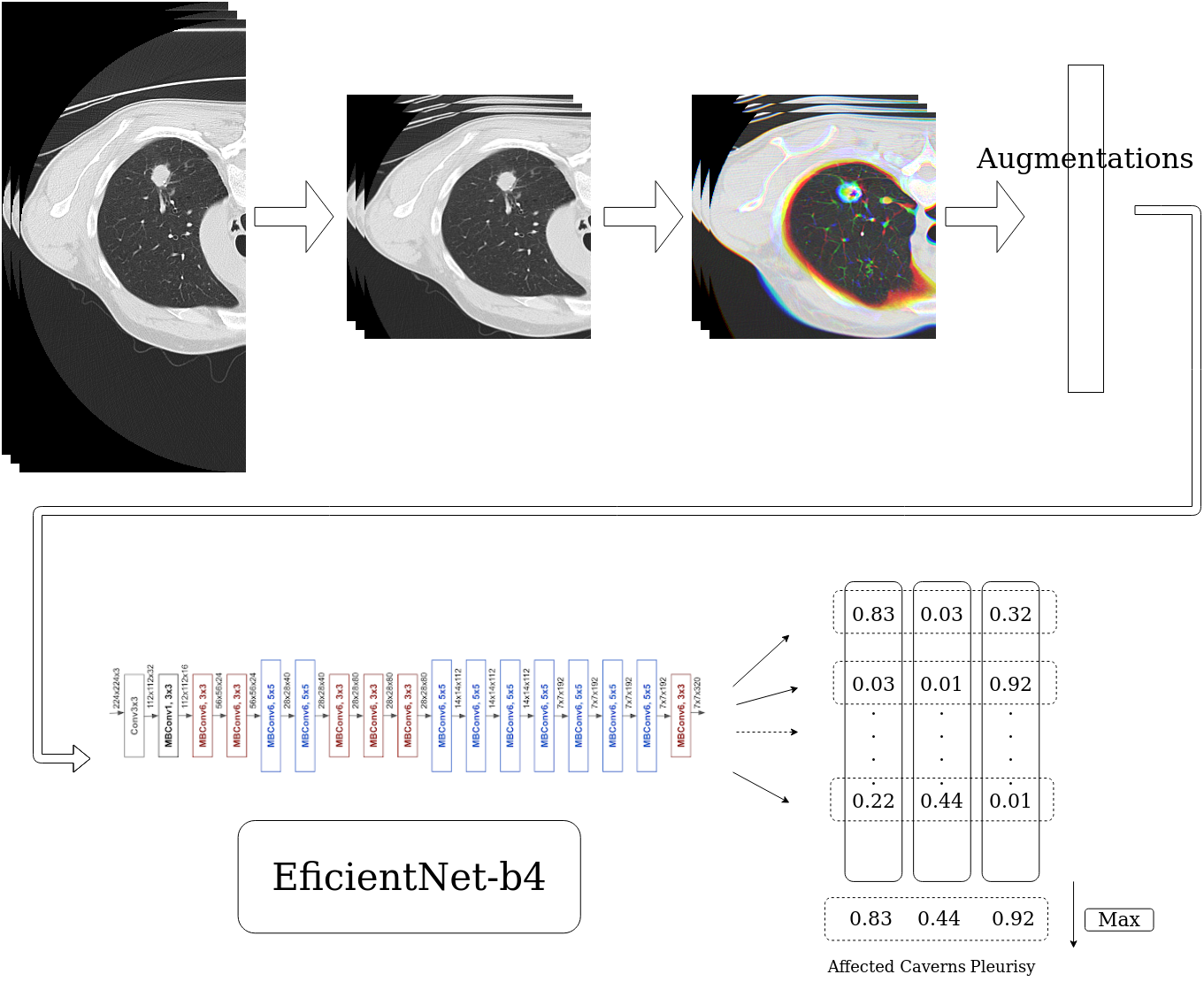}
\caption{The schematic of the slice approach (microvolumes). It follows a simple flow. The volumetric image, split by left/right side is cropped using a simple threshold based segmentation, then composed (in this case) to microvolumes, augmented and passed through the model. The output from all the slices of a side of a volume is then aggregated and the max per label selected to compose the final result for a side. }
\label{fig:}
\end{figure}
Micro-volumes and just repeating the image gave similar results on the test set (92.2\% and 92.4\% respectively), however the training on the "naive" case was done on 130 epochs on 3 GPUs with batch 56 $\times$ 3 whereas the "micro-volumes" case was done on 60 epochs on 2 GPUs with batch $56 \times 2$.
Therefore, the approach with the highest score on our hold out set was c) the simple one which also represented the highest of the submitted scores. Our second highest submitted model is represented by the approach in a) micro-volumes.

For training we used Nvidia RTX 2070 with 8GB of memory. We again, used the Apex library from NVidia to train using mixed precision and DistributedDataParallel with one process per GPU.

\section{Comparative results}\label{sec:results}

\subsection{Results on the competition test set}
 Table \ref{tab:my_label} summarizes the results obtained in the competition on the test set. Submissions were made only for the methods based on 2D projections and the ones based on predictions at slice level; as shown on the hold-out validation data, the attempts to use
the whole volume using 3D convolutions or fusing the information at slice level did not obtain good results and therefore were not used in the competition test phase.
\begin{table}[h]
    \centering
        \caption{Results reported on the test set, in the order of submission. The first two entries use 2d projections, while all the others make predictions at slice levels. (CE suffix represents models with CrossEntropyLoss and BCE models with BinaryCrossEntropyLoss}
    \begin{tabular}{c|c|c}
      Method   & mean AUC & min AUC \\
      \hline
       ResNet50Proj & 0.825 &0.766\\
       \hline
       PreProcProj & 0.793 & 0.703\\
       \hline
       NaiveInception & 0.860 & 0.772\\
       \hline
       ThresholdInception & 0.887 & 0.821\\
       \hline
       AttentionInception &0.853 & 0.788\\
       \hline
       TwoPicInception & 0.892 & 0.830\\
       \hline
       NaiveSliceCE & \textbf{0.924} & \textbf{0.885}\\
       \hline
       MicroVolSliceCE & 0.922 & 0.860\\
       \hline
       NaiveSliceBCE & 0.899 & 0.862\\
    \end{tabular}

    \label{tab:my_label}
\end{table}

\subsection{Discussion}
By comparing the results both on our hold-out validation set and on the test set, the following conclusions can be drawn.

\begin{itemize}
    \item As indicated by the low training accuracy, the approaches using the entire volumetric data as a whole corresponding to the segmented lungs (described in section \ref{sec:whole}), involving 3D convolutions or slice fusion, seem to be overwhelmed by the amount of parameters to fit and are not able to identify the lesions in cases where these are small or present only on a few slices of the CT, or either converge slowly.
    \item The 2D approaches based on projections computed over the segmented volume (described in section \ref{sec:proj}) give (unreasonable) good results, which indicates that simple (normalized) statistics like mean, maximum and standard deviation, when used together, are able to catch important information about the presence of lesions in lung CTs. The quality of segmentation of the lungs is of critical importance in this case, as a bad segmentation may introduce noise into the projections. After obtaining the set of 2D projections, data augmentation increased the generalization capability of the classifier.
    \item The best approach, surpassing significantly the ones based on 2D projections, exploits all the information present in the segmented volumetric lungs in a slice-wise manner. Instead of predicting the presence of the affection per CT, we predict it for each slice. To obtain the report back at lung level the probabilities over slices are aggregated by extracting the maximum. An important pre-processing step consisted in fixing the window and range levels to specific values used by radiologists when inspecting lung CTs.
    
\end{itemize}

\section{Conclusions}\label{sec:conclusions}
Volumetric images like CTs and MRIs provide rich information about the internal body structure, necessary in the diagnosis of many affections. With the advancements of neural networks, automatic diagnosis in volumetric images became possible at high precision, useful for prioritizing patients and assisting doctors in final decisions. After a thorough experimental analysis of various architectures, the current paper devised an approach able to produce highly accurate CT reports about the presence of tuberculosis related affections. The method, based on computing predictions at slice level, has, beside high accuracy in predicting lesion type, the advantage of offering more information in terms of localization of the lesions. With a current mean AUC score of 0.924 on test data, its performance can be increased if more data, capturing various cases, is provided in the training phase.

\section{Acknowledgements}
Our strong belief is that data science without domain knowledge can never reach its full potential. We would like to thank Mirela Iordache, an outstanding radiologist at the Regional Institute of Oncology in Iasi, for sharing her knowledge and valuable insights on the medical niche approached in the competition.

\bibliographystyle{unsrt}
\bibliography{biblio}

\end{document}